\documentclass[conference]{IEEEtran}
\usepackage{cite}

\ifCLASSINFOpdf
\else
\fi
\usepackage[cmex10]{amsmath}
\usepackage{mathrsfs}
\usepackage{amsthm}
\usepackage{accents}
\usepackage{statex}
\usepackage{graphicx}
\usepackage{enumerate}

\begin{document}

\newtheorem{Theorem}{Theorem}
\newtheorem{Lemma}[Theorem]{Lemma}

%
\title{Secret Key Generation from Correlated Sources and Secure Link}

\author{\IEEEauthorblockN{Daming Cao and Wei Kang}
\IEEEauthorblockA{School of Information Science and Engineering\\
Southeast University, Nanjing, P.R.China, 210096\\
Email:\{dmcao,wkang\}@seu.edu.cn}
}


%


\maketitle

\begin{abstract}
In this paper, we study the problem of secret key generation from both correlated sources and a secure channel. We obtain the optimal secret key rate in this problem and show that the optimal scheme is to conduct secret key generation and key distribution jointly, where every bit in the secret channel will yield more than one bit of secret key rate. This joint scheme is better than the separation-based scheme, where the secure channel is used for key distribution, and as a result, every bit in the secure channel can only provide one bit of secret key rate.
\end{abstract}



%
\IEEEpeerreviewmaketitle

\section{Introduction}
The problem of secret key generation was introduced by Ahlswede and Csisz{\'a}r \cite{ahlswede1993common} and Maurer \cite{maurer1993secret}, where two separate terminals, named Alice and Bob, observe the outcomes of a pair of correlated sources separately and want to generate a common secret key, which is concealed from an eavesdropper Eve, given that the terminals can communicate through a noiseless public channel which the eavesdropper has complete access to. In \cite{ahlswede1993common}, the secret key capacity of correlated sources was characterized when Alice and Bob are allowed to communicate once over a channel with unlimited capacity. The secrecy key capacity when there is a constraint on the rate of the public channel was found by Csisz{\'a}r and Narayan \cite{csiszar2000common} as a special case of their result when there is another terminal called the "helper" that is not interested in recovering the key but rather helping generating the secret key. In \cite{prabhakaran2008secrecy, prabhakaran2012secrecy,bassi2016secret}, the authors considered the case where Alice communicate over a noisy broadcast channel rather than a noiseless public channel, i.e., the wiretap channel.

In some applications, in addition to the public channel between Alice and Bob, there may exist a secure link between Alice and Bob as well. One example is in wireless sensor network, where the nodes wants to share a secret key to encrypt their communication. In this application, the frequency selectivity of the fading channel will create both public and secure channels between Alice and Bob. More specifically, in some frequency bands, the links from Alice to both Bob and Eve are of good qualities, which constitute the public channel, while in some other frequency bands, the link from Alice to Bob is of good quality, but the link from Alice to Eve is basically broken. These frequency bands can be viewed as  a secure channel. Another example comes from \cite{thai2014physical}, where Alice and Bob are nodes equipped with multiple antennas and they communicate with the help of multiple single-antenna relays employing the amplify-and-forward strategy. 
We assume that some relays are "nice but curious"\cite{lima2007random}, while the other relays are simply nice. Therefore, the links through the curious relays are public, while the links through the nice relays are secure. 

Motivated by the above applications, in this paper, we consider the problem of secret key generation where Alice, in addition to the public channel, has a secure channel to communicate with Bob. We explore the optimal strategy when the correlated sources, the public channel and the secure link  between Alice and Bob,  are all available. Under this scenario, one nature scheme is a separation-based scheme, where the public channel is used for key generation and the secure channel is used for key distribution. In this scheme, every bit in the secure channel can add one bit of secret key rate.
In this paper, we obtain the secret key capacity  and show that in the optimal scheme, the secure channel is used to conduct both the secret key generation and the key distribution jointly. In this joint scheme, every bit in the secure channel can yield more than one bit of the secret key rate, and therefore, it outperforms the separation-based scheme.

\begin{figure}
\centering\includegraphics[width=3.5in]{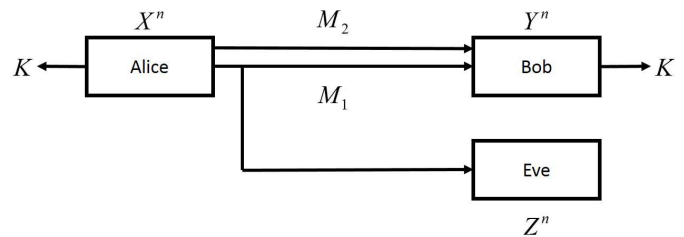}
\caption{Key Generation with Additional Secure Channel}\label{model}
\end{figure}

\section{System Model and Main Result}

Consider a network with three nodes, including a transmitter Alice, a receiver Bob and an eavesdropper Eve. We assume three discrete memoryless sources indicated by random variables $(X,Y,Z)$, defined in the alphabets $(\mathcal{X},\mathcal{Y},\mathcal{Z})$, respectively.  We assume that Alice and Bob observe the $n$-length source sequences $X^n$ and $Y^n$, respectively, and Eve observes $n$-length source sequence $Z^n$. 
A key generation code consists of two encoding functions $f_1$ and $f_2$ and two decoding functions $g_1$ and $g_2$, defined as follows:
\begin{align}
  f_1:&\mathcal{X}^n \mapsto \{1,2,\dots,L_1\}  \\
  f_2:&\mathcal{X}^n \mapsto \{1,2,\dots,L_2\}  \\
  g_1:& \mathcal{X}^n 
  \mapsto \{1,2,\dots,J\} \\
  g_2: &\mathcal{Y}^n \times \{1,2,\dots,L_1\}\times \{1,2,\dots,L_2\} 
  \mapsto \{1,2,\dots,J\}
\end{align}
Here, $M_1=f_1(X^n)$ is sent through the public channel and $M_2=f_2(X^n)$ is sent through the secure channel. Then the secret key $K$ is generated by Alice and Bob from the functions $g_1$ and $g_2$, respectively, which should agree with probability $1$ and be concealed from Eve.
The probability of error for the key generation code is defined as $P_e^{(n)}=\Pr\{g_1(X^n)\neq g_2(Y^n,M_1,M_2)\}$. 
A secret key rate $R_K$ with constraint communication rate pair $(R_1,R_2)$ is achievable if for any $\epsilon > 0$ there exists a key generation code such that
\begin{align}
  P_e^{(n)} &\leq \epsilon \\
  \frac{1}{n}I(K;Z^n,M_1) &\leq \epsilon \label{SMSC1}\\
  \frac{1}{n}\log J &\geq R_K-\epsilon\\
  \frac{1}{n}\log L_1 &\leq R_1+\epsilon\label{SMSC2}\\
  \frac{1}{n}\log L_2 &\leq R_2+\epsilon\label{SMSC3}
\end{align}
The main result of this paper is the following theorem.
\begin{Theorem}\label{mainth}
For given sources $(X^n,Y^n,Z^n)$, the rate triple $(R_K,R_1,R_2)$ is achievable if and only if
\begin{align}
  R_K&\le I(U;Y|T)-I(U;Z|T)+R_2\label{TH1}\\
  R_1+R_2 &\ge I(U;X|Y)  \\
  R_1&\ge I(T;X|Y) \label{TH11}
  \end{align}
  where random variables $(U,T,X,Y,Z)$ satisfy the following Markov chain
  \begin{align}
  T \rightarrow U &\rightarrow X \rightarrow (Y,Z)
\end{align}
\end{Theorem}

\section{The Converse}
We begin the proof of the converse with
\begin{align}\nonumber
n&R_K
  \leq  H(K)\nonumber\\
  &\leq H(K)-H(K|Y^n,M_1,M_2)+n\epsilon \label{ConK1}\\
   &=I(K;Y^n,M_1,M_2)+n\epsilon \nonumber\\
   &\leq  I(K;Y^n,M_1,M_2)-I(K;Z^n,M_1)+2n\epsilon \label{ConK11}\\
   &=I(K;Y^n,M_1)+I(K;M_2|Y^n,M_1)\nonumber\\
   &\quad-I(K,M_2;Z^n,M_1)-I(M_2;Z^n,M_1|K)+2n\epsilon \nonumber\\
   &=I(K,M_2;Y^n,M_1)+I(K;M_2|Y^n,M_1)-I(M_2;Y^n,M_1|K)\nonumber\\
   &\quad-I(K,M_2;Z^n,M_1)-I(M_2;Z^n,M_1|K)+2n\epsilon \nonumber\\
   &=I(K,M_2;Y^n,M_1)+I(K;M_2|Y^n,M_1)-I(M_2;Y^n,M_1,K)\nonumber\\
   &\quad-I(K,M_2;Z^n,M_1)-I(M_2;Z^n,M_1,K)+2n\epsilon \nonumber\\
   &=I(K,M_2;Y^n,M_1)-I(M_2;Y^n,M_1)\nonumber\\
   &\quad-I(K,M_2;Z^n,M_1)-I(M_2;Z^n,M_1,K)+2n\epsilon 
      \end{align}
   \begin{align}
   &=I(K,M_2;Y^n|M_1)-I(K,M_2;Z^n|M_1)\nonumber\\
   &\quad-I(M_2;Y^n,M_1)-I(M_2;Z^n,M_1,K)+2n\epsilon \nonumber\\
   &= I(K,M_2;Y^n|M_1)-I(K,M_2;Z^n|M_1)+H(M_2)\nonumber\\
   &\quad-I(M_2;Y^n,M_1)-H(M_2|Z^n,M_1,K)+2n\epsilon \nonumber\\
   &\leq I(K,M_2;Y^n|M_1)-I(K,M_2;Z^n|M_1)+nR_2+2n\epsilon
\end{align}
where (\ref{ConK1}) follows from Fano's inequality; (\ref{ConK11}) follows from the secrecy constraint in (\ref{SMSC1}). 

By applying the key identity \cite[Lemma 17.12]{csiszar2011information}, it follows that
\begin{align}
  I(K,M_2;Y^n|M_1)&-I(K,M_2;Z^n|M_1)\nonumber\\
  =& n[I(U;Y_J|T)-I(U;Z_J|T)]
\end{align}
where
\begin{align}
T&\triangleq (M_1,Y^{i-1},Z_{i+1}^n,J)\label{Tdef}\\
U&\triangleq (K,M_2,T)\label{Udef}
\end{align}
Here, $J$ is uniformly distributed on $\{1,\dots,n\}$ and independent of $(X^n,Y^n,Z^n)$. Since $K$, $M_1$ and $M_2$ is a function of $X^n$, the Markov Chain $T \rightarrow U \rightarrow X \rightarrow YZ $ is satisfied. Because of the fact that $(X^n,Y^n,Z^n)$ are i.i.d., we can replace $Y_J$ and $Z_J$ by $Y$ and $Z$. Then (\ref{TH1}) is proved.

Then we consider the sum rate $(R_1+R_2)$
\begin{align}
 n&(R_1+R_2)\nonumber\\
   &\geq H(M_1,M_2)\nonumber\\
   &\geq H(M_1,M_2|Y^n) \nonumber\\
   &\geq H(K,M_1,M_2|Y^n)-n\epsilon \label{ConSR1}\\
   &\geq H(K,M_1,M_2|Y^n)-H(K,M_1,M_2|X^n) -n\epsilon\nonumber \\
   &= I(K,M_1,M_2;X^n)-I(K,M_1,M_2;Y^n)-n\epsilon\nonumber\\
   &= \sum_{i=1}^n[I(K,M_1,M_2;X_i|X_{i+1}^n,Y^{i-1})\nonumber\\
   &-I(K,M_1,M_2;Y_i|X_{i+1}^n,Y^{i-1})]-n\epsilon \label{ConSR2}\\
   &= \sum_{i=1}^n[I(K,M_1,M_2,X_{i+1}^n,Y^{i-1};X_i)\nonumber\\
   &-I(K,M_1,M_2,X_{i+1}^n,Y^{i-1};Y_i)] -n\epsilon\label{ConSR3}\\
   &= \sum_{i=1}^n[I(K,M_1,M_2,X_{i+1}^n,Y^{i-1},Z_{i+1}^n;X_i)\nonumber\\
   &-I(K,M_1,M_2,X_{i+1}^n,Y^{i-1},Z_{i+1}^n;Y_i)] -n\epsilon\label{ConSR4}\\
   &= \sum_{i=1}^n[I(K,M_1,M_2,X_{i+1}^n,Y^{i-1},Z_{i+1}^n;X_i|Y_i)-n\epsilon\label{ConSR41}\\
   &\geq \sum_{i=1}^n[I(K,M_1,M_2,Y^{i-1},Z_{i+1}^n;X_i)\nonumber\\
   &-I(K,M_1,M_2,Y^{i-1},Z_{i+1}^n;Y_i)]-n\epsilon\label{ConSR5}
\end{align}
where (\ref{ConSR1}) follows from Fano's inequality; (\ref{ConSR2}) follows from the key identity \cite[Lemma 17.12]{csiszar2011information}; (\ref{ConSR3}) follows because $(X^n,Y^n,Z^n)$ are i.i.d., (\ref{ConSR4}) follows from the Markov Chain $Z_{i+1}^n \rightarrow (K,M_1,M_2,X_{i+1}^n,Y^{i-1}) \rightarrow (X_i,Y_i) $; (\ref{ConSR41}) follows from the Markov Chain $Y_i\rightarrow X_i\rightarrow (K,M_1,M_2,X_{i+1}^n,Y^{i-1},Z_{i+1}^n)$; (\ref{ConSR5}) follows from the Markov Chain $X_{i+1}^n \rightarrow (K,M_1,M_2,Y^{i-1},Z_{i+1}^n,X_i) \rightarrow Y_i $. Then we have
\begin{align}
  n(R_1+R_2) &\geq  n[I(U;X_J)-I(U;Y_J)] \nonumber\\
   &= nI(U;X_J|Y_J)\nonumber\\
   &= nI(U;X|Y)\nonumber
\end{align}
with $U$ as defined in (\ref{Udef}) and $J$ is uniform on $\{1,2,\dots,n\}$.

Finally, for the public rate $R_1$, we have
\begin{align}
  nR_1 &\geq H(M_1)\nonumber\\
  &\geq H(M_1|Y^n) \nonumber\\
   &\geq H(M_1|Y^n)-H(M_1|X^n) \nonumber\\
   &= I(M_1;X^n)-I(M_1;Y^n)\nonumber
\end{align}
By the similar argument as in the sum rate derivation, we have 
\begin{align}
  nR_1 &\geq \sum_{i=1}^n[I(M_1Y^{i-1}Z_{i+1}^n;X_i)-I(M_1Y^{i-1}Z_{i+1}^n;Y_i)]  \nonumber\\
   &\geq  n[I(T;X_J)-I(T;Y_J)] \nonumber\\
   &= nI(T;X_J|Y_J)\nonumber\\
   &= nI(T;X|Y)
\end{align}
where $T$ as defined in (\ref{Tdef}) and $J$ is uniform on $\{1,2,\dots,n\}$,  which concludes the proof of the converse.

\section{The Achievability}
The achievability scheme that we propose in this paper can be viewed as a combination of the scheme in secret key generation from correlated sources \cite{csiszar2000common}, consisting of a codebook of the superposition structure, and secret key distribution through secret channel.
The rate of the public channel and the secure channel are used for the transmission of the following:
\begin{enumerate}
\item the inner code $T^n$.
\item the outer code $U^n$.
\item key distribution.
\end{enumerate}

Our proposed scheme follows the principles below:
\begin{itemize}
\item The public channel is used to transmit the inner code $T^n$. Since the rate of the inner code is less than the rate of the public channel, then the leftover rate of the public channel will be used to transmit the outer code $U^n$.
\item  The secure channel is used to transmit the outer code $U^n$. Since there is still extra rate leftover in the secure channel, then the leftover rate of the secure channel can be used for key distribution.
\item The public channel can not be used for key distribution and the secure channel can not be used to transmit the inner code $T^n$.
\end{itemize}

We show that the proposed scheme achieves the rate in Theorem \ref{mainth} and is thus, optimal. The details of the proof can be found in the appendix.

%
%

\section{Discussions}

We first elaborate on the principles of the proposed scheme in the previous section. Since distributing key through public channel will compromise the confidentiality of the key, the public channel can not be used for key distribution. Regarding the principle where the secure channel does not transmit the inner code, from the proof of the achievability in the appendix, we will find that the conditional distribution $P_{T|X}$ for the inner code satisfies
\begin{align}
I(T;Y)\le I(T;Z)
\end{align}
which means,
Eve can obtain extra knowledge about $T^n$ from $Z^n$ than what Bob gets from $Y^n$. This extra knowledge means that though part of $T^n$ is transmitted  through the secure channel, it is not secure because Eve can infer this secure part of $T^n$ from her observation of $Z^n$.

Regarding the principle that for the rate of the secure channel, the transmission of the outer code gets higher priority than key distribution, we explain by comparing our proposed scheme with the separation-based scheme, where the public channel is used for key generation and the secure channel is used for key distribution. The secret key rate of the separation-based scheme is
\begin{align}
  R_K&\le I(U;Y|T)-I(U;Z|T)+R_2\\
  R_1&\ge I(U;X|Y)
  \end{align}
  where random variables $(U,T,X,Y,Z)$ satisfy the following Markov chain
  \begin{align}
  T \rightarrow U &\rightarrow X \rightarrow (Y,Z)
\end{align}
%
%
%
%
Comparing the above secret key rate to the secret key rate in Theorem \ref{mainth}, we find that the expressions for the secret key rate $R_K$ are the same for both schemes, but the constraints on the public and secure rates are  stricter for this separation-based scheme, which results in a smaller value of the term $I(U;Y|T)-I(U;Z|T)$, and therefore a smaller secret key rate in this separation based scheme.
The reason is that in the separation-based scheme, the secure channel is only used for key distribution. Therefore, every bit of the secure channel rate gives one bit of secret key rate. However, for the proposed joint scheme, which is optimal, the secure channel is used to transmit the outer code. Every bit through the secure channel for the outer code remains secure to the eavesdropper, and therefore yields one bit of secret key rate. In addition, one bit of the outer code through the secure channel helps in the scheme of key generation by delivering one more bit to Bob, which gives a larger secret key rate. We conclude that every bit that we increase on $R_2$ can contribute in two different aspects in the joint scheme and yield more than one bit of the increase in secret key. Hence, the proposed joint scheme outperforms the separation based scheme. This explains why, for the rate of the secure channel, the transmission of the outer code gets higher priority than key distribution.

Based on the discussion of the above paragraph, we conclude that when fixing $R_1$, every bit increased in $R_2$ will provide more than one bit of secrecy key rate. Next, consider the case where the sum rate $R_1+R_2$ is fixed. We note that increasing $R_2$ will improve the optimal $R_K$ because we can always use the secure channel as the public channel, which will keep $R_K$ the same. Then the optimal scheme will outperform this simple scheme and yield a larger secret key rate. If we look into the expressions in Theorem \ref{mainth}, we can find that increasing $R_2$ (and therefore decreasing $R_1$), on one hand, will increase the last term $R_2$ in the expression of $R_K$ in (\ref{TH1}), but on the other hand, will decrease the first two terms $I(U;Y|T)-I(U;Z|T)$ of $R_K$ in (\ref{TH1}) by tighten the constraint on the distribution in (\ref{TH11}). Therefore, we conclude that every bit that we move from $R_1$ to $R_2$ will provide no more than one bit of increase on the optimal secret key rate $R_K$.

\section{Conclusion}
In this paper, we studied the problem of key generation from both correlated sources and a secure channel. We found the secret key capacity when there are rate constraints on both channels, by providing the proofs of both achievability and the converse. We showed that the optimal scheme is to use the secure channel for secret key generation and key distribution jointly, in which case every bit in the secure channel can yield more than one bit of secret key. This optimal scheme is better than the separation-based scheme where the public channel is used for secret key generation and the secure channel is used for key distribution, in which one bit in secure channel can only give one bit of secret key.


\appendix
\section{The Achievability}
Consider a given distribution
\begin{equation}
  p(tuxyz)=p(ut)p(x|u)p(yz|x)
\end{equation}

\textbf{Case 1}: If $R_1\geq I(U:X|Y)$

We use a separate scheme as follows. Alice use the public channel to achieve a secrecy rate $I(U;Y|T)-I(U;Z|T)$ by the scheme in secret key generation, and use the secure channel to distribute another secret key with rate $R_2$. Combine two secret keys together,  the secrecy rate in (\ref{TH1}) is achievable.

\textbf{Case 2}: If $R_1\leq I(U:X|Y)$.

In this case, we assume that $I(T;Y)\le I(T;Z)$. The reason is that if $I(T;Y)>I(T;Z)$, we define a pair of new random variable $(U',T')$ such that $U'=(U,T)$ and $T'=\emptyset$. Then we have $I(T';Y)\le I(T';Z)$. From the three inequalities in the main theorem, we have
\begin{align}
R_K&\le I(U;Y|T)-I(U;Z|T)+R_2\nonumber\\
&\le I(U';Y|T')-I(U';Z|T')+R_2\\
  R_1+R_2 &\ge I(U;X|Y)=I(U';X|Y)  \\
  R_1&\ge I(T;X|Y)\ge I(T';X|Y)
\end{align}
The above derivation shows that for every pair $(U,T)$ with Markov chain $ T' \rightarrow U' \rightarrow X $ and  $I(T;Y)>I(T;Z)$, we can find another pair $(U',T')$ with Markov chain $ T \rightarrow U \rightarrow X $ and  $I(T';Y)\le I(T';Z)$ such that the rate region  with $(U',T')$ is achievable implies that the rate region with $(U,T)$ is also achievable. Therefore, we only need to consider the case where $I(T;Y)\le I(T;Z)$. 

We define the following notations
\begin{align}
    R_{11} &= I(T:X|Y) \\
    R_{12} &= R_1-R_{11} \\
    R_{21} &= I(U;X|Y,T)-R_{12} \\
    R_{22} &= R_2-R_{21} \\
    R_T &= I(T;X)\\
    R_U &= I(U;X|T)\\
    R_{K_1} &= I(U;Y|T)-I(U;Z|T)
\end{align}

\emph{Codebook generation}:
\begin{enumerate}
  \item Randomly and independently generation $2^{nR_T}$ sequences $t^n(s_1)$, $s_1\in[1:2^{nR_{T}}]$ according to distribution $p(t)$ and randomly and independently partition them into $2^{nR_{11}}$ bins with bin indices $B_1(m_{11})$, $m_{11}\in[1:2^{nR_{11}}]$.
  \item For each codeword $t^n(s_1)$, randomly and independently generation $2^{nR_U}$ sequences $u^n(s_1,s_2)$, $s_2\in[1:2^{nR_{U}}]$ according to $p(u|t)$, and randomly and independently partition them into $2^{n(R_{12}+R_{21})}$ bins with bin indices $B_2(s_1,m_{12},m_{21})$, $m_{12}\in[1:2^{nR_{12}}]$ and $m_{21}\in[1:2^{nR_{21}}]$. Randomly and independently partition the sequences in each nonempty bin $B_2(s_1,m_{12},m_{21})$ into $2^{nR_{K_1}}$ bins with bin indices  $B_2(s_1,m_{12},m_{21},k_1)$, $k_1\in[1:2^{nR_{K_1}}]$.
\end{enumerate}

\emph{Encoding}:
\begin{enumerate}
  \item Given a source sequence $X^n$, find the index $s_1$ such that $(t^n(s_1),X^n)$ is jointly typical with respect to the joint distribution $p(t,x)$, i.e.,  $(t^n(s_1),X^n)\in \mathcal{T}^n_{[TX]_\delta}$. If there is no such index or there are more than one such indices, then randomly select $s_1$ from $[1:2^{nR_{T}}]$, where the probability of such event is arbitrarily small if $R_T > I(T;X)$. 
  Let $B_1(m_{11})$ be the bin index of $t^n(s_1)$. 
  \item Then find the index $s_2$ such that $(t^n(s_1),u^n(s_1,s_2),X^n)\in \mathcal{T}^n_{[TUX]_\delta}$.  If there is no such index or there are more than one such indices, then random select $s_2$ from $[1:2^{nR_{U}}]$,  where the probability of such event is arbitrarily small if $R_U > I(U;X|T)$. Let $B_2(s_1,m_{12},m_{21},k_1)$ be the bin index of $u^n(s_1,s_2)$.
  \item Randomly choose $k_2\in[1;2^{nR_{22}}]$.
  \item Alice sends $M_1=(m_{11},m_{12})$ and $M_2=(m_{21},k_2)$ to Bob through the public and secure channel, respectively.
  \item Alice chooses $K=(k_1,k_2,m_{21})$ as the secret key.
\end{enumerate}

\emph{Decoding}:   Upon receiving $(M_1,M_2)$, Bob decodes the secret key as following:
\begin{enumerate}
  \item Find the unique sequence $t^n(\hat{s}_1)\in B_1(m_{11})$ such that $(t^n(\hat{s}_1),y^n)\in \mathcal{T}^n_{[TY]_\delta}$. And the probability that there is no such sequence or there are more than one such sequences is arbitrarily small if
      \begin{equation}
        R_T-R_{11} < I(T;Y)
      \end{equation}
  \item Find the index $\hat{s}_2\in B_2(\hat{s}_1,m_{12},m_{21})$  such that $(t^n(\hat{s}_1),u^n(\hat{s}_1,\hat{s}_2),y^n)\in \mathcal{T}^n_{[TUY]_\delta}$. Let $B_2(\hat{s}_1,m_{12},m_{21},\hat{k}_1)$ be the bin index of $u^n(\hat{s}_1,\hat{s}_2)$. The probability that there is no such sequence or there are more than one such sequences is arbitrarily small if
      \begin{equation}
        R_U-R_{12}-R_{21} < I(U;Y|T)
      \end{equation}
  \item Let $\hat{K}=(\hat{k}_1,k_2,m_{21})$
\end{enumerate}

\emph{Key leakage rate}: We consider the key leakage rate averaged over the random coding scheme $C$ as described above. We begin with
\begin{align}
  H&(K|Z^n,M_1) \nonumber\\
   &= H(K,U^n,T^n|Z^n,M_1)-H(U^n,T^n|Z^n,K,M_1) \nonumber \\
   &= H(U^n,T^n|Z^n,M_1)+H(K|U^n,T^n,Z^n,M_1)\nonumber\\
   &\quad-H(U^n,T^n|Z^n,K,M_1) \nonumber\\
   &= H(T^n|Z^n,M_1)+H(U^n|Z^n,M_1,T^n)+H(K_2)\nonumber\\
   &\quad-H(T^n|K,Z^n,M_1)-H(U^n|K,Z^n,M_1,T^n)\label{k2}\\
   &= I(T^n;K|Z^n,M_1)+H(U^n|Z^n,T^n)+H(K_2)\nonumber\\
   &\quad-I(U^n;M_1|Z^n,T^n)-H(U^n|K,Z^n,M_1,T^n)\nonumber\\
   &\ge H(U^n|Z^n,T^n)+H(K_2)+H(M_1|Z^n,T^n,U^n)\nonumber\\
   &\quad-H(M_1|Z^n)-H(U^n|K,Z^n,M_1,T^n)\nonumber\\
   &\geq H(U^n|Z^n,T^n)+H(K_2)-H(M_1)\nonumber\\
   &\quad -H(U^n|K,Z^n,M_1,T^n)\label{w1}\\
   &\ge H(U^n|Z^n,T^n)+nR_{22}-nR_1\nonumber\\
   &\quad-H(U^n|K_1,M_{21},Z^n,M_1,T^n)\label{final}
\end{align}
where (\ref{k2}) follows because $K=(K_1,K_2, M_{21})$, in which $K_2$ is independent of $(U^n,T^n,Z^n,M_1)$ and $K_1$ and $M_{21}$ are functions of $U^n$; (\ref{w1}) follows because $M_1=(M_{11},M_{12})$ is a function of $(T^n,U^n)$; (\ref{final}) follows because $K_2$ is uniformly distributed, $H(M_1)$ is upper bounded by $nR_1$, and $K=(K_1,M_{21})$.


For the first term in (\ref{final}), we have
\begin{align}
 \frac{1}{n}H(U^n|Z^n,T^n)
  &=\frac{1}{n}[H(U^n|T^n)-I(Z^n;U^n|T^n)\nonumber\\
  &= R_U-I(Z;U|T)
\end{align}


For the fourth term, similar to \cite[Lemma 22.3]{el2011network}, we can bound it in the following lemma.
\begin{Lemma} If $R_{K_1}< H(U|Z,T)-H(U|Y,T)-2\delta(\epsilon)$
\begin{align}
  \lim_{n\rightarrow \inf} \sup& \frac{1}{n} H(U^n|K_1,M_{21},Z^n,M_1,T^n) \nonumber\\
  &\leq R_U-R_{K_1}-R_{21}-R_1-I(Z;U|T)+\delta(\epsilon)
\end{align}
\end{Lemma}

From the above derivation,  we have
\begin{align}
  \frac{1}{n}I(K;Z^n,M_1)&=\frac{1}{n} [H(K)-H(K|Z^n,M_1)] \nonumber\\
  &\leq \frac{1}{n}H(K)-[R_{K_1}+R_{21}+R_{22}-\delta(\epsilon)]\nonumber\\
  &\le \delta(\epsilon)
\end{align}
which concludes the proof of the achievability.

\bibliography{cdmbibliography}
\bibliographystyle{IEEEtran}

\end{document}